\documentclass[twocolumn,aps,prl] {revtex4} 
\usepackage{graphicx}
\begin{document}
\pagestyle{empty} 
\title{Fluid squeeze-out between solids with rough surfaces}
\author{B. Lorenz$^{1,2}$ and B.N.J. Persson$^1$}
\affiliation{$^1$ IFF, FZ J\"ulich, D-52425 J\"ulich, Germany}
\affiliation{$^2$ IFAS, RWTH Aachen University, D-52074 Aachen, Germany}

\begin{abstract}
We study the fluid squeeze-out from the interface between an elastic solid with a flat surface
and a rigid solid with a randomly rough surface. 
As an application we discuss fluid squeeze-out between a tire tread block and a road surface. 
Some implications for the leakage of seals are discussed, and experimental data are presented to
test the theory.
\end{abstract}
\maketitle


{\bf 1. Introduction}

Contact mechanics between solid surfaces is the basis for understanding many
tribology processes\cite{Bowden,Johnson,BookP,Isra,Sealing,Capillary.adhesion,P33} 
such as friction, adhesion, wear and sealing. The two most important
properties in contact mechanics are the area of real contact and the interfacial separation between the
solid surfaces. 
For non-adhesive contact and small squeezing pressure, 
the average interfacial separation depends 
logarithmically on the squeezing pressure\cite{P4,P3},
and the (projected) contact area depends linearly on the squeezing pressure\cite{P1}.
Here we study how the (average) interfacial separation depends on time when two elastic
solids with rough surfaces are squeezed together in a fluid. In particular, we calculate
the time necessary to squeeze-out the fluid from the contact regions between the solids.
As an application we discuss fluid squeeze-out between a tire tread block and a road surface.  
Some implications for the leakage of seals are discussed, and experimental data are presented to
test the theory.

\begin{figure}
\includegraphics[width=0.35\textwidth,angle=0]{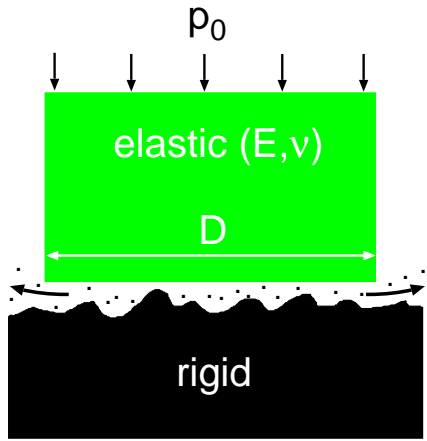}
\caption{\label{pic}
The upper solid, a cylindrical block
with the diameter $D=2R$, the elastic modulus $E$ and the Poisson ratio $\nu$, 
is squeezed in a fluid against a rigid substrate. The bottom surface of the cylinder
is perfectly flat and the substrate surface 
randomly rough with the root-mean-square roughness amplitude
$h_{\rm rms}$.} 
\end{figure}

\vskip 0.3cm
{\bf 2. Squeeze-out: large separation}

Consider an elastic solid with a flat surface squeezed in a fluid against a rigid 
solid with a randomly rough surface, see Fig. \ref{pic}.
The fluid is assumed to be Newtonian with the viscosity $\eta$. The upper solid is a cylindrical block
with the radius $R$, the elastic modulus $E$ and the Poisson ratio $\nu$. 
The bottom surface of the cylinder
is perfectly flat, and the substrate randomly rough with the root-mean-square roughness amplitude
$h_{\rm rms}$. Here we focus first on the the simplest possible 
situation which can be studied analytically, 
where the (macroscopic or locally averaged) pressure distribution in the fluid gives rise to negligible 
deformations of the bottom surface of the elastic block. This requires
that the amplitude $\delta u$ of the (fluid-induced) elastic deformations is much smaller than  
$h_{\rm rms}$. Since $\delta u$ is typically of order (or smaller than)  $\approx p_0 R/ E^*$, 
where $E^*=E/(1-\nu^2)$ and
$p_0$ is the pressure applied to the upper surface of the cylinder block, we 
get the condition $h_{\rm rms} >> p_0 R /E^*$. For elastically stiff materials with
$E^* \approx 10^{11} \ {\rm Pa}$ and for $R = 1 \ {\rm cm}$ we get $p_0 R /E^* < 0.1 \ {\rm \mu m}$ if
$p_0 < 1 \ {\rm MPa}$. For tread rubber $E^* \approx 10 \ {\rm MPa}$ and if $p_0 \approx 0.3 \ {\rm MPa}$
we get $p_0 R /E^* \approx 0.3 \ {\rm mm}$ which is smaller than the root-mean-square roughness of many 
asphalt road surfaces. In many applications a thin rubber film, coating a hard solid, is in contact
with an elastically hard countersurface. If the linear size of the contact region is large compared to the rubber film thickness,
this geometry will strongly suppress the (fluid-induced) deformations of the rubber film on the length scale of the linear size of the 
nominal contact area, and most of the (non-uniform) deformations of the rubber film is due to the
interaction with the substrate asperities. 

We first develop a theory which should be accurate for large enough interfacial separation, 
e.g., corresponding to the early phase of the squeeze-out process.
We assume that the longest wavelength roughness component, $\lambda_0$, is small compared to the linear size
$R$ of the (apparent) contact region. In this case we can speak about locally averaged (over surface
areas with linear dimension of order $\lambda_0$) quantities. 

Neglecting inertia effects, the squeeze-out is determined by (see, e.g., Ref. \cite{BookP})
$${d \bar u\over dt} \approx -{2\bar u^3\bar p_{\rm fluid} (t) \over 3 \eta R^2},\eqno(1)$$
where $\bar p_{\rm fluid} (t)$ is the (average) fluid pressure, 
and $\bar u$ the (locally averaged) interfacial separation. 
If $p_0$ is the applied pressure acting on the top surface of the cylinder block,
we have
$$\bar p_{\rm fluid} (t) = p_0 - p_{\rm cont}(t), \eqno(2)$$
where $p_{\rm cont}$ is the asperity contact pressure. 
We will first assume that the pressure $p_0$ is so small that for all times 
$u >> h_{\rm rms}$. In this case we can use the asymptotic relation\cite{P4}
$$p_{\rm cont} \approx \beta  
E^* {\rm exp}\left ( - {\bar u \over u_0}\right ).\eqno(3)$$
where $u_0 = h_{\rm rms}/\alpha$. The parameters $\alpha$ and $\beta$ depends on the fractal properties 
of the rough surface\cite{P4}. 

From (3) we get
$${d \bar u \over dt}  \approx - {u_0 \over p_{\rm cont}} {d p_{\rm cont} \over dt}. \eqno(4)$$
Using (4) and (2) we get from (1):
$${d p_{\rm cont} \over dt} \approx {2\bar u^3 (p_{\rm cont}(t)) \over 3 \eta R^2 u_0}p_{\rm cont}\left (p_0-p_{\rm cont}\right ),\eqno(5)$$
For long times $p_{\rm cont} \approx p_0$ and  
we can approximate (5) with
$${d p_{\rm cont} \over dt} \approx {2 \bar u^3 (p_0) \over 3 \eta R^2 u_0} p_0 (p_0-p_{\rm cont}).$$
Integrating this equation gives
$$p_{\rm cont} (t) \approx p_0 - \left [p_0-p_{\rm cont}(0)\right ] {\rm exp} 
\left (-\left ({\bar u (p_0)\over h_{\rm rms}}\right )^3 {t\over \tau} \right )$$
where
$$\tau = {3\eta R^2 u_0 \over 2 h^3_{\rm rms} p_0} = 
{3 \eta R^2  \over 2 \alpha h^2_{\rm rms} p_0}.\eqno(6)$$
Using (3) this gives
$$\bar u \approx  u_\infty + \left ( 1-{p_{\rm cont}(0)\over p_0} \right ) 
u_0 {\rm exp} \left ( -\left ({\bar u (p_0)\over h_{\rm rms}}\right )^3 {t\over \tau} \right )$$
where $u_\infty = u_0 {\rm log}(\beta E^*/p_0)$.
Thus, $\bar u(t)$ will approach the equilibrium separation $u_\infty$ in an exponential way, and
we can define the squeeze-out time as the time to reach, say, $1.01 u_\infty$. 
For flat surfaces, within continuum mechanics, the film thickness approach zero as $t\rightarrow \infty$ as
$\bar u \sim t^{-1/2}$. Thus in this case  
it is not possible to define a meaningful fluid squeeze-out time.

Let us measure distance $\bar u$ in units of $h_{\rm rms}$, pressure in units of $p_0$ and
time in units of $\tau$ (eq. (6)). 
In these units (5) takes the form 
$${d p_{\rm cont} \over dt} \approx \bar u^3 p_{\rm cont}\left (1-p_{\rm cont}\right ),\eqno(7)$$
In the same units (3) takes the form
$$p_{\rm cont} \approx \kappa^{-1} {\rm exp} \left ( - \alpha \bar u \right ).\eqno(8)$$
where
$$\kappa = {p_0 \over \beta E^*} \eqno(9)$$

\begin{figure}
\includegraphics[width=0.45\textwidth,angle=0]{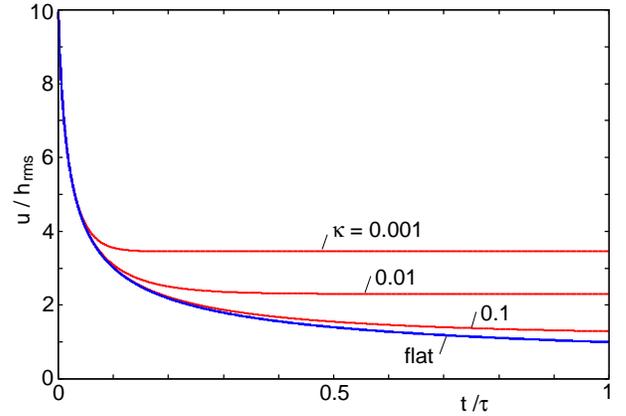}
\caption{\label{time.u}
The interfacial separation
(or film thickness) $\bar u$ (in units of $h_{\rm rms}$) as a function of the
squeeze-time $t$ (in units of $\tau$) for several 
values of the parameter $\kappa$. For each
squeeze-pressure, the upper (red) lines are the result for the rough
surfaces while the lower (blue) lines are for flat surfaces.} 
\end{figure}

In Fig. \ref{time.u} we show the interfacial separation
(or film thickness) $\bar u$ (in units of $h_{\rm rms}$) as a function of the
squeeze-time $t$ (in units of $\tau$) for several 
values of the parameter $\kappa$. For each
$\kappa$ value, the upper (red) lines are the result for the rough
surfaces while the lower (blue) lines are for flat surfaces. 
In the calculation we have used $\alpha = 0.5$ and $\beta = 1$.

\begin{figure}
\includegraphics[width=0.45\textwidth,angle=0]{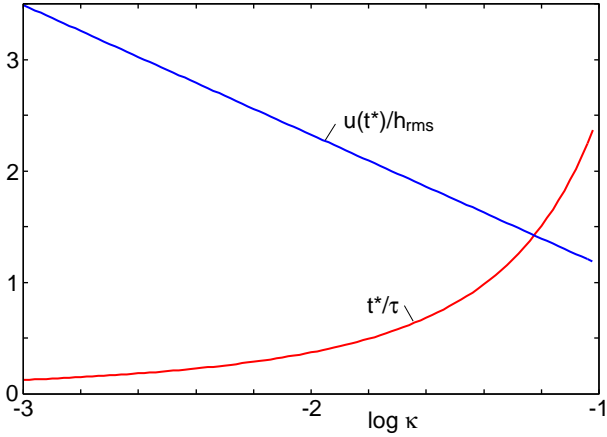}
\caption{\label{barSigma.time}
The fluid squeeze-out time $t^*$ (in units of $\tau$) and the final interfacial separation
(or film thickness), $\bar u(t^*)$ (in units of $h_{\rm rms}$), as a function of the
logarithm (with 10 as basis) of the parameter $\kappa$. We define $t^*$ so that $\bar u(t^*)=1.01 \bar u(\infty)$.} 
\end{figure}

In Fig. \ref{barSigma.time} we show
the fluid squeeze-out time $t^*$ (in units of $\tau$) and the final interfacial separation
(or film thickness), $\bar u(t^*)$ (in units of $h_{\rm rms}$), as a function of the
parameter $\kappa$. We define $t^*$ so that $\bar u(t^*)=1.01 \bar u(\infty)$.

At high enough squeezing pressures, the interfacial separation after long enough times will
be smaller than $h_{\rm rms}$, and
the asymptotic relation (3) will no longer hold. In this case the relation
$p_{\rm cont} (\bar u )$ can be calculated using the equations given in Ref. \cite{Sealing} 
(see also Appendix A). Substituting (2) in (1) and measuring pressure in units of
$p_0$, separation in units of $h_{\rm rms}$ and time in units of $\tau$ one obtain 
$${d \bar u\over dt} \approx -\alpha^{-1} \bar u^3 (1-p_{\rm cont}),\eqno(10)$$
where $\alpha = h_{\rm rms}/u_0$.
This equation together with the relation $p_{\rm cont} (\bar u )$ constitute
two equations for two unknown ($\bar u$ and $p_{\rm cont}$) which are easily solved
by numerical integration. In what follows we refer to the theory
presented above as the average-separation theory.

\vskip 0.3cm
{\bf 3. Squeeze-out: general theory}

We now present a general theory of squeeze-out, which is accurate for small separation and which
reduces to the result presented in Sec. 2 for large separation.
The theory presented below is based on a recently developed 
theory of the leak-rate of (static) seals\cite{subm}.
We assume again that the longest wavelength roughness component, $\lambda_0$, is small compared to the linear size
$R$ of the (apparent) contact region. In this case we can speak about locally averaged (over surface
areas with linear dimension of order $\lambda_0$) quantities. 
Let ${\bf J} ({\bf x},t)$ 
be the (locally averaged) 2D-fluid flow vector which satisfies the continuity equation
$$\nabla \cdot {\bf J} +{\partial \bar u \over \partial t}=0,\eqno(11)$$
where $\bar u({\bf x},t)$ is the (locally averaged) surface separation or, equivalently, 
the 2D-fluid density (fluid volume per unit area). 
Here and in what follows 
$${\bf J} = (J_x,J_y), \ \ \ \ \  \nabla = (\partial_x, \partial_y) \eqno(12)$$ 
are 2D vectors.
In Ref. \cite{subm} we have shown that within an effective medium approach
$${\bf J} = - \sigma_{\rm eff} \nabla p_{\rm fluid},\eqno(13)$$
where $p_{\rm fluid}({\bf x},t)$ is the (locally averaged) fluid pressure and
where the effective conductivity $\sigma_{\rm eff} (p_{\rm cont})$ depends on the (locally averaged)
contact pressure $p_{\rm cont}({\bf x},t)$. Note that when inertia effects are negligible
$$\int d^2x \left [  p_{\rm cont}({\bf x},t)+ p_{\rm fluid}({\bf x},t)\right ] = F_{\rm N}(t)\eqno(14)$$
is the applied normal load. The function $\sigma_{\rm eff} (p_{\rm cont})$ can be calculated from
the surface roughness power spectrum and the (effective) 
elastic modulus as described in Ref. \cite{subm}. Substituting (12) and (13) in (11) gives
$$\nabla \cdot \left [\sigma_{\rm eff} \nabla p_{\rm fluid}\right ] 
= {\partial \bar u \over \partial t}. \eqno(15)$$
Eq. (15) together with the relation $p_{\rm cont}(\bar u)$
and the (standard) expression relating the macroscopic 
deformation $\bar u({\bf x},t)$ to the local pressure
$p_0({\bf x})=p_{\rm cont}({\bf x},t)+ p_{\rm fluid}({\bf x},t)$ constitutes 
three equations for the three unknown $p_{\rm cont}$, $p_{\rm fluid}$
and $\bar u$. In addition one need the effective medium expression 
for $\sigma_{\rm eff} (p_{\rm cont})$, and the
``boundary condition'' (14) must be satisfied. Here we will not study the most general problem but we focus on
the limiting case discussed above where the macroscopic deformations of the solid walls can be neglected.
In this case $\bar u$ and $p_{\rm cont}$ will only depend on time. As a result $\sigma_{\rm eff} (p_{\rm cont})$
will only depend on time. Thus, (15) reduces to
$$\sigma_{\rm eff} \nabla^2 p_{\rm fluid}
= {d \bar u \over dt}.\eqno(16)$$
Since the right hand side only depends on time, 
$$p_{\rm fluid} ({\bf x},t)= 2 \bar p_{\rm fluid} (t) \left [1-\left (r/R \right )^2 \right ]\eqno(17)$$
where $\bar p_{\rm fluid} (t)$ is the average (nominal) fluid pressure in the nominal contact region.
Substituting (17) in (16) gives 
$${8\sigma_{\rm eff}\over R^2}  \bar p_{\rm fluid} = - {d \bar u \over dt}.\eqno(18)$$
Eq. (14) takes the form
$$p_{\rm cont} + \bar p_{\rm fluid} = p_0\eqno(19)$$
Using (4) and (19) in (18) gives
$${d p_{\rm con}\over dt} \approx {8 \sigma_{\rm eff} (p_{\rm cont}) \over R^2 u_0} p_{\rm cont} (p_0-p_{\rm cont}).\eqno(20)$$
From this equation one obtain $p_{\rm cont}(t)$, and using (3) and (19) 
one can then calculate $\bar u(t)$ and $p_{\rm fluid}(t)$.
As shown in Appendix A, when $p_{\rm cont} \rightarrow 0$, $\bar u \rightarrow \infty$ and
$\sigma_{\rm eff} \rightarrow \bar u^3 /12\eta$, which is an exact result to leading order in $h_{\rm rms}/\bar u$. 
Substituting $\sigma_{\rm eff} = \bar u^3/12 \eta$ in (20) gives (5). Thus, in the limit of small pressures $p_0$, 
the present treatment reduces to the average-separation theory of Sec. 2, which is exact 
when the average separation $\bar u$ between the surfaces is large
(which is the case for all times if the pressures $p_0$ is small). We will refer to $\sigma_{\rm eff} = \bar u^3/12 \eta$ as the
average-separation expression for $\sigma_{\rm eff}$.

Let us study the squeeze-out for long times. For long times $p_{\rm cont} \approx p_0$ and  
we can approximate (20) with
$${d p_{\rm con}\over dt} \approx {8 \sigma_{\rm eff} (p_0) \over R^2 u_0} p_0 (p_0-p_{\rm cont}).\eqno(21)$$
Integrating this equation gives
$$p_{\rm cont} (t) \approx p_0 - \left [p_0-p_{\rm cont}(0)\right ] {\rm exp} \left (-{8\sigma_{\rm eff}(p_0) p_0\over R^2 u_0} t\right )$$
so that $p_{\rm cont} (t)$ approach $p_0$ (and $\bar u(t)$ approach $u_\infty$) in an exponential way, 
just as for the simpler model studied in Sec. 2

If we measure pressure in units of $p_0$, separation $\bar u$ in 
units of $h_{\rm rms}$, and time in units of
$\tau$, Eq. (20) takes the form
$${d p_{\rm con}\over dt} =  \bar \sigma_{\rm eff} p_{\rm cont} (1-p_{\rm cont}), \eqno(22)$$
where 
$$\bar \sigma_{\rm eff} = 12 \eta \sigma_{\rm eff} /h_{\rm rms}^3.\eqno(23)$$
The relation between $p_{\rm cont}$ and $\bar u$ is given by (3).

At high enough squeezing pressures, the interfacial separation after long enough times will
be smaller than $h_{\rm rms}$, and
the asymptotic relation (3) no longer hold. In this case the relation
$p_{\rm cont} (\bar u )$ can be calculated using the equations given in Ref. \cite{Sealing} 
(see also Appendix A). Substituting (19) in (18) and measuring pressure in units of
$p_0$, separation in units of $h_{\rm rms}$ and time in units of $\tau$ one obtain 
$${d \bar u\over dt} = -\alpha^{-1} \bar \sigma_{\rm eff} (1-p_{\rm cont})\eqno(24)$$
This equation together with the relations $p_{\rm cont} (\bar u )$ and $\sigma_{\rm eff} (p_{\rm cont})$ constitute
three equations for three unknown ($\bar u$, $p_{\rm cont}$ and $\sigma_{\rm eff}$) which are easily solved
by numerical integration. In the critical junction theory which we will use below $\sigma_{\rm eff} = (\alpha u_1 (\zeta_{\rm c}))^3/12\eta$,
where the separation $u_1 (\zeta_{\rm c})$ in defined in Ref. \cite{subm} (see also Appendix A) and where $\alpha < 1$ is a number
of order unity (see Ref. \cite{subm}).

\begin{figure}
\includegraphics[width=0.45\textwidth,angle=0]{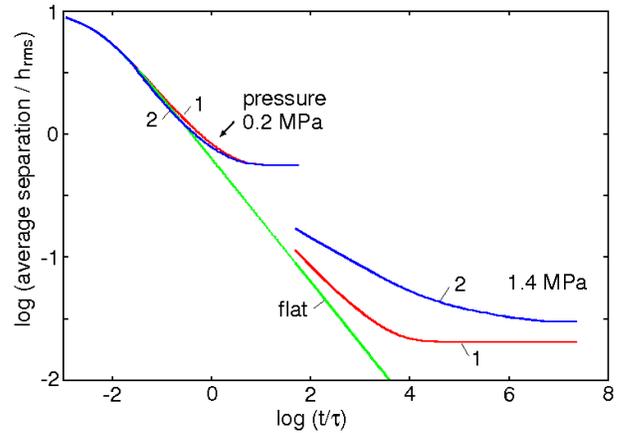}
\caption{\label{logascale.time.separation1}
The interfacial separation $\bar u (t)$ as a function of the
squeeze-time $t$ for rough copper surface (log-log scale with 10 as basis). 
Curves {\bf 1} and {\bf 2} are the predictions
using the average-separation theory and using the critical-junction theory, respectively. 
Results are also shown for a flat surface.
The rubber block is assumed to be cylindrical with the radius $R=1.5 \ {\rm cm}$.
For a copper surface with the root-mean-square roughness $0.12 \ {\rm mm}$. 
The squeezing pressure $p_0=0.2 \ {\rm MPa}$ (upper curves) and $1.4 \ {\rm MPa}$.
The elastic modulus $E=2.3  \ {\rm MPa}$.} 
\end{figure}

\begin{figure}
\includegraphics[width=0.45\textwidth,angle=0]{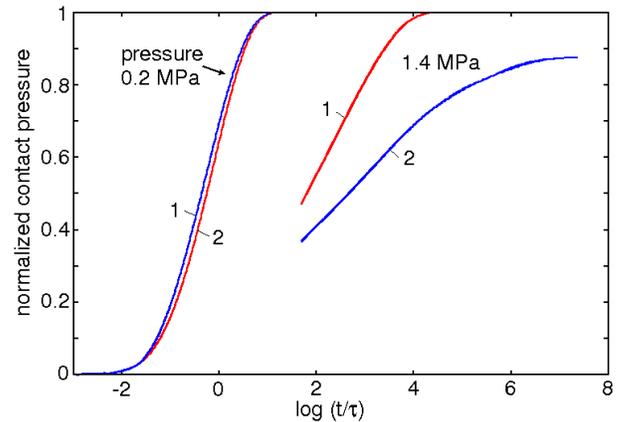}
\caption{\label{logascale.time.contactPressure}
The normalized contact pressure $p_{\rm con} (t)/p_0$ as a function of the
logarithm (with 10 as basis) of the squeeze-time $t$ for rough copper surface.
Curves {\bf 1} and {\bf 2} are the predictions
using the average-separation and using the critical-junction theory, respectively. 
The rubber block is assumed to be cylindrical with the radius $R=1.5 \ {\rm cm}$.
For a copper surface with the root-mean-square roughness $0.12 \ {\rm mm}$. 
The squeezing pressure $p_0=0.2 \ {\rm MPa}$ (upper curves) and $1.4 \ {\rm MPa}$.
The elastic modulus $E=2.3  \ {\rm MPa}$.} 
\end{figure}

Fig. \ref{logascale.time.separation1}
shows the calculated interfacial separation $\bar u (t)$ as a function of the
squeeze-time $t$ for a silicon rubber block (elastic modulus $E=2.3  \ {\rm MPa}$) 
squeezed against a 
rough copper surface (log-log scale with 10 as basis), with the power spectrum given in Ref. \cite{HeatTransfer}. 
Curves {\bf 1} and {\bf 2} are the theory predictions
using (10) and (24), respectively.
In (24) we have used $\sigma_{\rm eff}$ as calculated using the critical-junction theory
described in Ref. \cite{subm}, which gives nearly the same result as the effective medium theory described in the same reference.
Results are also shown for a flat substrate surface.
The rubber block is assumed to be cylindrical with the radius $R=1.5 \ {\rm cm}$ and the
surface of the copper block has the root-mean-square roughness $0.12 \ {\rm mm}$. 
The squeezing pressure $p_0=0.2 \ {\rm MPa}$ (upper curves) and 
$1.4 \ {\rm MPa}$ (lower curves), corresponding to $\kappa=p_0/\beta E^*$ 
$\approx 1.5$ and 10.3, respectively.
Note that for the pressure $p_0=1.4 \ {\rm MPa}$, 
after long enough time the area of real contact, $A$, percolate
(i.e., $A/A_0 > 0.5$, see Ref. \cite{subm}),
and there is no fluid leak channel at the interface\cite{strictly}. As a result,  
when the contact area percolate the
fluid is confined at the interface and is not able to leak-out. Thus, even after very long time 
the interfacial separation is larger than would be expected in the absence of trapped
or confined fluid (e.g., for dry contact),
where no part of the load would be carried by the fluid. 

Fig. \ref{logascale.time.contactPressure}
shows the normalized contact pressure $p_{\rm con} (t)/p_0$ as a function of the
logarithm of the squeeze-time $t$ for the same system as in Fig. \ref{logascale.time.separation1}.
Curves {\bf 1} and {\bf 2} are the predictions
using (10) and (24), respectively. 
Note that for the pressure $p_0=1.4 \ {\rm MPa}$, even after very long enough time $p_{\rm con} <0.9 p_0$. 
This is again a consequence of the fact that the non-contact area does not percolate 
and fluid is confined at the interface, and even after very long time, more than $10\%$ of the 
external load is carried by the confined fluid.

Following Ref. \cite{MSc} the analysis presented above may be extended to include
the fluid-pressure induced elastic deformation of the solid surfaces at the interface.
It is also easy to include the dependency of the
fluid viscosity on the local pressure or local surface separation. The former is important for
elastically hard solids (high pressures), e.g., steel, and the latter ``confinement effect''
even for elastically soft solids (low pressures)
when the fluid film thickness becomes of order a few nanometer or less\cite{Shi}.

\begin{figure}
\includegraphics[width=0.35\textwidth,angle=0]{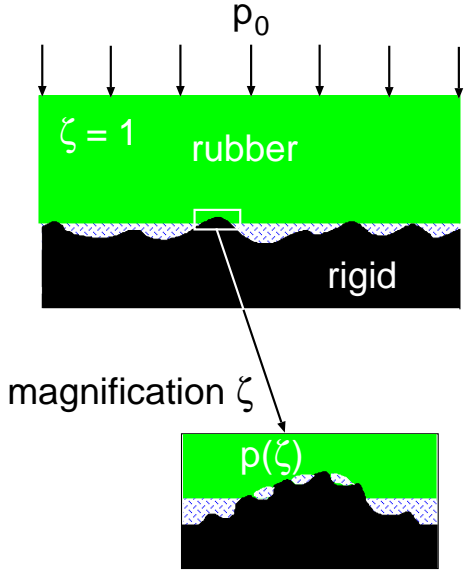}
\caption{\label{twoMagnification}
A rubber block squeezed against a randomly rough substrate in a fluid. The (average) interfacial separation 
$\bar u(t)$ decreases with time $t$ due to squeeze-out of the fluid. The time dependence of $\bar u$ depends mainly on the
long-wavelength roughness components.
Nevertheless, the short wavelength roughness components will affect the squeeze-out in asperity contact regions (observed at the
magnification $\zeta$), which is relevant for, e.g., rubber friction. This can be studied using the same theory applied
to an asperity contact region by replacing the external pressure $p_0$ with the local squeezing 
pressure $p(\zeta )$ at the asperity (see text for details).  
} 
\end{figure}

Finally, let us note the following: The interfacial separation is 
usually mainly determined by the most long-wavelength surface roughness, which is observed close
to the lowest magnification $\zeta \approx 1$; the shortest wavelength
(small amplitude) roughness has almost no influence on $\bar u$. However, it is also
of great interest, e.g., in the context of tire friction on wet road surface (see Sec. 4), 
to study how the fluid is squeezed out from the (apparent) asperity contact regions
observed at higher magnification $\zeta$, see Fig. \ref{twoMagnification}. 
The theory developed above can be applied to this
case too. Thus, let us study the squeeze-out of fluid from the apparent asperity
contact regions observed at the magnification $\zeta$. At this magnification no surface roughness
with wavelength below $L/\zeta$ can be observed. However, when the magnification is increased
one observe shorter wavelength roughness which will influence the fluid squeeze-out, and which may even
result in sealed-off, trapped fluid. We can apply the theory above to
study the squeeze out of fluid from the asperity contact regions observed at the magnification
$\zeta$ by using instead of the external pressure $p_0$, the local squeezing
pressure $p(\zeta) = p_0 A_0/A(\zeta)$, where $A(\zeta)$ is the (apparent) contact area
observed at the magnification $\zeta$. The surface roughness in the contact regions is given by the
surface power spectrum $C(q)$ for $q > \zeta /L$. With these modifications we can use 
the theory above to calculate the squeeze-out of fluid from the apparent asperity contact regions
observed at the magnification $\zeta$. One complication is, however, that the fluid is squeezed-out
from an asperity contact region into the surrounding, and the fluid pressure in the surrounding may be higher
than the external pressure (which we have taken as our reference pressure in the study above) existing
outside the nominal contact region observed at the lowest magnification $\zeta = 1$. As a result,
in order to study the squeeze-out from the asperity contact regions observed at the magnification
$\zeta'$, one must first study the squeeze-out from the asperity contact regions observed at lower magnification
$1 < \zeta < \zeta'$. 
We will not develop this theory here, but we believe a similar approach as that used to describe
mixed lubrication for flat on flat in Ref. \cite{MSc} may be applied to the present problem.

\begin{figure}
\includegraphics[width=0.45\textwidth,angle=0]{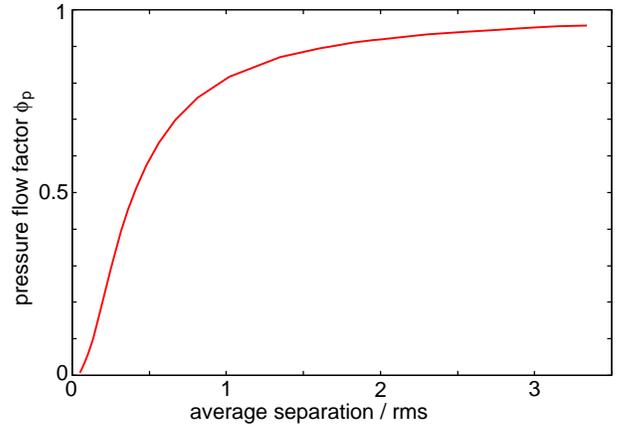}
\caption{\label{1average.separation.2effective.medium.flow.factor}
The pressure flow factor $\phi_p$, calculated using the Bruggeman effective medium theory (see Ref. \cite{subm}), 
as a function of the average surface separation
$\bar u$ for a rough copper surface with the root-mean-square roughness $0.12 \ {\rm mm}$. 
The squeezing pressure $p_0=1.0 \ {\rm MPa}$ and the elastic modulus $E=2.3  \ {\rm MPa}$.} 
\end{figure}

\vskip 0.3cm
{\bf 4. Pressure flow factor}

The theory presented above can be used to calculate the pressure flow factor $\phi_p$ first introduced
by Patir and Cheng\cite{Patir}. 
This quantity is defined so that the (locally averaged) flow current ${\bf J}$ associated with fluid flow at the interface between two 
stationary solids with rough surfaces,
is given by
$${\bf J} = - {1 \over 12 \eta} \bar u^3 \phi_p \nabla  p_{\rm fluid}$$
where $\bar u$ and $p_{\rm fluid}$ are the locally averaged interfacial separation and the locally averaged (nominal) fluid pressure, respectively.
Using (13) and (23) we get
$$\phi_p = 12 \eta \sigma_{\rm eff} / \bar u^3 = \bar \sigma_{\rm eff} (h_{\rm rms} / \bar u )^3$$
In Fig. \ref{1average.separation.2effective.medium.flow.factor} we show the
pressure flow factor $\phi_p$, 
calculated using the Bruggeman effective medium theory (see Ref. \cite{subm}), as a function of the average surface separation
$\bar u$ for a rough copper surface with the root-mean-square roughness $0.12 \ {\rm mm}$. 
The squeezing pressure $p_0=1 \ {\rm MPa}$ and the elastic modulus $E=2.3  \ {\rm MPa}$. 
Note that $\phi_p$ vanish already for a non-zero $\bar u$. This is due to the percolation of the contact area. Note also
that $\phi_p \rightarrow 1$ as $\bar u \rightarrow \infty$. This result is expected because for
large separation the surface roughness should have a negligible influence on the fluid flow. 

The pressure flow factor $\phi_p$ is usually determined by calculating (numerically) 
the fluid flow in small interfacial units\cite{Patir}. Most studies
neglect the elastic deformations of the solid walls, and only take into account surface roughness over
two decades (or less) in length scale. The present treatment includes elastic deformation and can easily take into account roughness
on arbitrary many decades in length scale. In Ref. \cite{PSeal} we have shown how one can generalize the treatment above to obtain the flow
factor (now a matrix) for surfaces with random roughness with anisotropic statistical properties.

\vskip 0.3cm
{\bf 5. Application to tire on wet road}

\begin{figure}
\includegraphics[width=0.45\textwidth,angle=0]{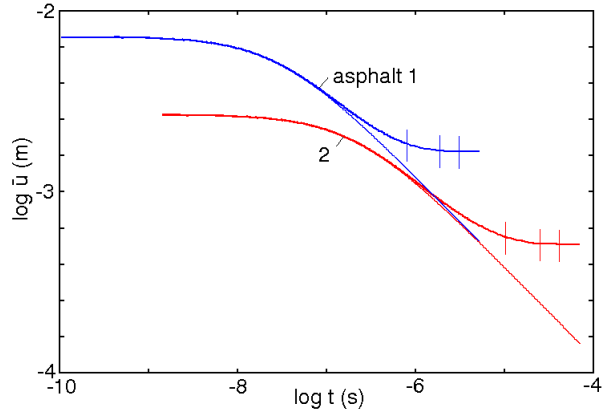}
\caption{\label{logascale.time.separation}
The interfacial separation $\bar u (t)$ as a function of the
squeeze-time $t$ for two asphalt road surfaces (log-log scale with 10 as basis).
The tread block is assumed to be cylindrical with the radius $R=2 \ {\rm cm}$.
For two asphalt road surfaces with the root-mean-square roughness $0.72 \ {\rm mm}$
(surface {\bf 1}) and $0.24 \ {\rm mm}$ (surface {\bf 2}). 
The nominal squeezing pressure $p_0=0.2 \ {\rm MPa}$ and the elastic modulus $E=10  \ {\rm MPa}$. 
The vertical lines denote (from left to right) the 
time when $\bar u(t)/\bar u(\infty) = 1.1$, $1.01$ and $1.001$. The thin lines
denote the fluid film thickness for perfectly flat surfaces. The calculations use the average-separation
expression for $\sigma_{\rm eff}$.} 
\end{figure}

\begin{figure}
\includegraphics[width=0.45\textwidth,angle=0]{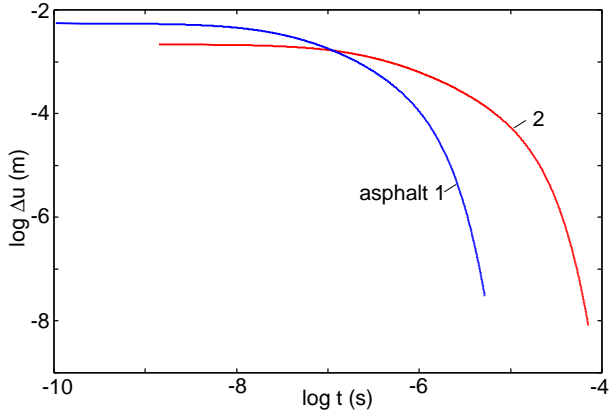}
\caption{\label{log.scale.time.separation.minus.asymphtot}
The difference $\Delta u = \bar u(t)- \bar u (\infty)$ as a function of the
squeeze-time $t$ for two asphalt road surfaces (log-log scale with 10 as basis).
The tread block is assumed to be cylindrical with the radius $R=2 \ {\rm cm}$.
For two asphalt road surfaces with the root-mean-square roughness $0.72 \ {\rm mm}$
(surface {\bf 1}) and $0.24 \ {\rm mm}$ (surface {\bf 2}). 
The squeezing pressure $p_0=0.2 \ {\rm MPa}$ and the elastic modulus $E=10  \ {\rm MPa}$.} 
\end{figure}

As an application, consider a tire tread block squeezed against 
asphalt road surfaces in water. In Fig. \ref{logascale.time.separation} and
\ref{log.scale.time.separation.minus.asymphtot} we show the time dependence of the 
interfacial separation $\bar u$, and the difference $\Delta u = \bar u(t)- \bar u (\infty)$,
respectively. In the calculation we have used the theory of Sec. 2, which is valid in the present
case where (for all times) $\bar u > 2 h_{\rm rms}$ (see below). 
We show results for two road surfaces, with the root-mean-square roughness
$0.72 \ {\rm mm}$ (surface {\bf 1}) and $0.24 \ {\rm mm}$ (surface {\bf 2}). We assumed 
the squeezing pressure $p_0=0.2 \ {\rm MPa}$ and the elastic modulus $E=10  \ {\rm MPa}$.
In the calculation we have used the surface roughness
power spectrum obtained from the measured surface topographies. 
Note that for the smoother surface the squeeze-out time is roughly one decade longer
than for the rougher asphalt road surface. In order for the water to have a negligible
influence on the hysteresis contribution to the 
friction, the water layer in the road-rubber contact regions must be
smaller than $\sim 1 \ {\rm \mu m}$. For the smoother road surface
${\rm 2}$ it takes about $\sim 3 \times 10^{-5} \ {\rm s}$ to reach
$\Delta u = 1 \ {\rm \mu m}$. 
If the tire rolling
velocity is $30 \ {\rm m/s}$ and the length of the tire foot-print $0.1 \ {\rm m}$,
then a tread block spend about $3\times 10^{-3} \ {\rm s}$ in contact with the road.
Thus, from the calculation above one may conclude that accounting just for the viscosity of the
fluid (water) one expect (during rolling) almost complete fluid squeeze-out from the tread-block road contact area,
during most of the time the tread block spends in the footprint. However, at the start of
raining after a long time of dry road condition, the water will be mixed with
contamination particles (e.g., small rubber and road wear particles), and the effective viscosity of the
mixture may be much larger than for pure water. In this case the squeeze-out may be incomplete,
which could result in viscous hydroplaning during braking. 
In addition, even for pure water, regions of sealed off (trapped) 
fluid may appear at the interface at high enough magnification,
which will reduce the hysteresis contribution to the tire-road friction\cite{Nature}.
We note that during braking at small slip (below the maximum in the $\mu$-slip curve)
the tread block does not slip until close to the exit of the tire-road footprint,
and the discussion above should therefore be valid for this case too.

It is interesting to compare the result above with the squeeze-out time due to inertia
(but neglecting the viscosity). The time-dependence of the squeeze-out for flat surfaces
is given by (see Ref. \cite{frog})
$$u(t)=u(0){\rm exp}(-t/\tau'), $$
where
$$\tau' \approx \left ({D^2 \rho \over 64 p_0}\right )^{1/2}.$$
Thus, the time it takes to reduce the film thickness from $u(0)$ to a thickness of order
$h_{\rm rms}$ is
$$t^* \approx \tau' {\rm log} [u(0)/h_{\rm rms}].$$
If $u(0) = 1 \ {\rm cm}$ and $h_{\rm rms} = 0.3 \ {\rm mm}$ we get 
$t^* \approx \tau' {\rm log} (30) \approx 3.4 \tau'$. With $D \approx 2 \ {\rm cm}$, 
$p_0 \approx 0.3 \ {\rm MPa}$ and $\rho \approx 10^3 \ {\rm kg/m^3}$ we get $\tau' 
\approx 1.4 \times 10^{-4} \ {\rm s}$ and $t^* \approx 5\times 10^{-4} \ {\rm s}$.   

\vskip 0.3cm
{\bf 6. Leakage of seals}

\begin{figure}
\includegraphics[width=0.25\textwidth,angle=0]{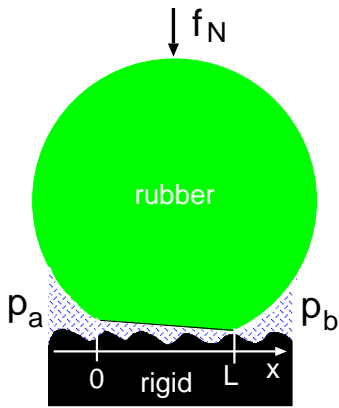}
\caption{\label{ballsqueezed}
Cross-section of rubber O-ring squeezed against a rigid, randomly rough surface in a fluid.
The fluid pressure is higher at $x\approx 0$ than for $x\approx L$, i.e.,
$p_{\rm a} > p_{\rm b}$, which result in fluid flow at the
interface, from the right to the left. When $p_{\rm a}$ is comparable to the
nominal squeezing pressure $f_{\rm N}/L$ (where $f_{\rm N}$ is the squeezing 
force per unit length of the cylinder), 
the elastic solid wall will deform and tilt
as indicated in the figure.
} 
\end{figure}

In the calculation of the leak-rate of seals presented in Ref. \cite{subm} we neglected the
influence of the fluid pressure on the contact mechanics. This is a good approximation
as long as the squeezing pressure $p_0$ is much higher than the 
fluid pressure $p_{\rm fluid}$, which was the case in the experiments presented in Ref. \cite{subm}.
However, in many practical situations it is not a good approximation to neglect the influence of the
fluid pressure on the contact mechanics. Since the fluid pressure is higher
on the fluid entrance side than on the fluid exit side, one expect the elastic wall to deform
and tilt relative to the average substrate surface plane, see Fig. \ref{ballsqueezed}.
Here we show how one can include the fluid
pressure when calculating the leak rate of seals. For simplicity we focus on the simplest
case where the fluid pressure only depends on one coordinate $x$, as would be the case for most
seal applications, e.g., rubber O-ring seals, see Fig. \ref{ballsqueezed}. 
In this case, for a stationary situation (15) takes the form
$${d\over dx}\left (\sigma_{\rm eff}(p_{\rm cont}(x)){d\over dx} p_{\rm fluid}(x)\right ) = 0$$ 
or
$$\sigma_{\rm eff}(p_{\rm cont}(x)){d\over dx} p_{\rm fluid}(x) = B$$ 
where $B$ is a constant. From this equation we get
$$p_{\rm fluid}(x) = A+B\int_0^x dx' \ \sigma^{-1}_{\rm eff}(p_{\rm cont}(x'))\eqno(25)$$
where $A$ is a constant.
If the fluid pressure for $x<0$ (high pressure side) is denoted by $p_{\rm a}$, and for $x>L$ (low pressure side) by $p_{\rm b}$, 
then using that $p_{\rm fluid}(0) = p_{\rm a}$ and $p_{\rm fluid}(L) = p_{\rm b}$ we can
determine the constants $A$ and $B$ in (25) and get $A=p_{\rm a}$ and
$$B= {p_{\rm b}-p_{\rm a} 
\over \int_0^L dx' \ \sigma^{-1}_{\rm eff}(p_{\rm cont}(x'))}\eqno(26)$$
Substituting these results in (25) gives
$$p_{\rm fluid}(x) = p_{\rm a}+(p_{\rm b}-p_{\rm a}){\int_0^x dx' \ \sigma^{-1}_{\rm eff}(p_{\rm cont}(x'))
\over \int_0^L dx' \ \sigma^{-1}_{\rm eff}(p_{\rm cont}(x'))} \eqno(27)$$

The rubber O-ring is squeezed against the substrate by the normal
force per unit radial length, $f_{\mathrm{N}}$, see Fig. \ref{ballsqueezed}.
In the contact region between the cylinder and
the substrate occur a nominal (locally averaged) pressure:
$$p_0(x)=p_{\mathrm{cont}}(x) + p_{\mathrm{fluid}}(x). \eqno(28)$$
We consider a stationary case so that 
$$\int_{-\infty}^{\infty} d x \ p_0(x) = {\frac{f_{\mathrm{N}}}{L}}.\eqno(29)$$
The elastic deformation field\cite{Johnson} 
$$\bar u(x)=u_c+{\frac{x^2}{2R}} -{\frac{2}{\pi E^*}}\int_{-\infty}^{\infty}
dx^{\prime }\ p_0 (x^{\prime }) {\rm log} \left | {\frac{x-x^{\prime }}{%
x^{\prime }}}\right |. \eqno(30)$$
Equations (27), (28) and (30),
together with the equation determining 
the relation between $p_{\rm cont}$ and $\bar u$,
represent 4 equations for the 4 unknown variables $p_0$, $p_{\mathrm{cont}}$, $p_{\mathrm{fluid}}$ 
and $\bar u$. In addition the pressure $p_0(x)$ must satisfy the normalization condition (29)
which determines the parameter $u_c$ in (30). 

The leak-rate of the seal is given by $\dot Q = J_x L_y$ where $L_y$ is width of the seal
(e.g., the circumstance of the seal for a rubber O-ring). Using (13) we get
$$\dot Q = -L_y  \sigma_{\rm eff} (p_{\rm cont}(x)) {d\over dx} p_{\rm fluid}(x) = -L_y B$$
Using (26) this gives
$$\dot Q = {L_y (p_{\rm a}-p_{\rm b}) \over 
\int_0^L dx' \ \sigma^{-1}_{\rm eff}(p_{\rm cont}(x'))} \eqno(31)$$
If $p_{\rm cont}$ is constant this gives
$$\dot Q = {L_y\over L_x} \sigma_{\rm eff} (p_{\rm a}-p_{\rm b})\eqno(32)$$
where we now denote $L=L_x$. Eq. (32) agree with the result presented in Ref. \cite{subm}.

For elastically soft materials like rubber the calculation of the leak-rate presented above can be simplified because
a small change in the interfacial separation will have a negligible influence on the nominal stress distribution in the
nominal contact area. Thus we can consider $p_0(x)$ as a given fixed function obtained by squeezing the 
elastic solid against a flat surface in the absence of the fluid. Using (27) and (28) we get
$$p_{\rm cont}(x) = p_0(x)-p_{\rm a}+(p_{\rm a}-p_{\rm b}){\int_0^x dx' \ \sigma^{-1}_{\rm eff}(p_{\rm cont}(x'))
\over \int_0^L dx' \ \sigma^{-1}_{\rm eff}(p_{\rm cont}(x'))} \eqno(33)$$
This equation can be iterated to obtain the solution $p_{\rm cont}(x)$. If the interfacial separation $\bar u > 2 h_{\rm rms}$
we can obtain the interfacial separation $\bar u$ from $p_{\rm cont}$ using
$$\bar u =  u_0 {\rm log} (\beta E^* / p_{\rm cont}),$$
but in general the relation between $\bar u$ and $p_{\rm cont}$ must be calculated from (A4).

\begin{figure}
\includegraphics[width=0.22\textwidth,angle=0.0]{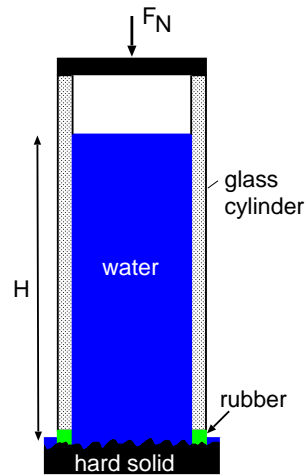}
\caption{\label{tube.water}
Experimental set-up for measuring the leak-rate of seals.
A glass (or PMMA) cylinder with a rubber ring attached to one end is squeezed against
a hard substrate with well-defined surface roughness. The cylinder is filled with 
water, and the leak-rate of the water at the rubber-countersurface
is detected by the change in the height of the water in the cylinder. 
}
\end{figure}

\vskip 0.3cm
{\bf 7. Experimental} 

We have performed a very simple 
experiment to test the theory presented in Sec. 6. 
In Fig.~\ref{tube.water} we show our 
set-up for measuring the leak-rate of seals.
A glass (or PMMA) cylinder with a rubber ring (with rectangular cross-section)
attached to one end is squeezed against
a hard substrate with well-defined surface roughness. The cylinder is filled with 
water, and the leak-rate of the fluid at the rubber-countersurface
is detected by the change in the height of the fluid in the cylinder. In this case
the pressure difference $\Delta p = p_{\rm a}-p_{\rm b} = \rho g H$, where $g$ is the gravitation
constant, $\rho$ the fluid density and $H$ the height of the fluid column. With $H\approx 1 \ {\rm m}$
we get typically $\Delta p \approx 0.01 \ {\rm MPa}$.
In the present study we use a rubber ring with the Young's elastic modulus $E=2.3 \ {\rm MPa}$, and with the inner and outer diameter
$4 \ {\rm cm}$ and $6 \ {\rm cm}$, respectively, and the height $0.5 \ {\rm cm}$. 
The rubber ring was made from a silicon elastomer (PDMS) prepared using
a two-component kit (Sylgard 184) purchased from Dow Corning (Midland, MI). The kit consist of a base 
(vinyl-terminated polydimethylsiloxane) and a curing agent (methylhydrosiloxane-dimethylsiloxane copolymer) 
with a suitable catalyst. From these two components we prepared a mixture 10:1 (base/cross linker) in weight.
The mixture was degassed to remove the trapped air induced by stirring from the 
mixing process and then poured into casts. The bottom of these casts was made from glass to obtain 
smooth surfaces. The samples were cured in an oven at $80 ^\circ {\rm C}$ for 12 h.

We have used a sand-blasted PMMA as substrate. 
The root-mean-square roughness of the surface is $34 \ {\rm \mu m}$.
In Ref. \cite{subm} 
we show the height probability distribution $P(h)$ and the
power spectrum $C(q)$ of the PMMA surface. 
 
\vskip 0.3cm
{\bf 8. Experimental results and comparison with theory}

In earlier studies we have performed experiments with the external load was 
so large that the condition $p_0>> \Delta p$ was satisfied, which is necessary in order to be able to
neglect the influence on the contact mechanics from the fluid pressure at the rubber-countersurface\cite{LorenzEPL,subm}.
However, here we are interested in the situation where the fluid pressure is comparable to the nominal squeezing pressure.
Thus the normal load is $18.5 \ {\rm N}$ giving the nominal squeezing pressure $p_0 = 11.8 \ {\rm kPa}$. Using a water
column with height $H=1.2 \ {\rm m}$ gives the fluid pressure $p_{\rm a}-p_{\rm b} = 11.8 \ {\rm kPa}$ at the bottom of the
fluid column.

\begin{figure}
\includegraphics[width=0.45\textwidth,angle=0]{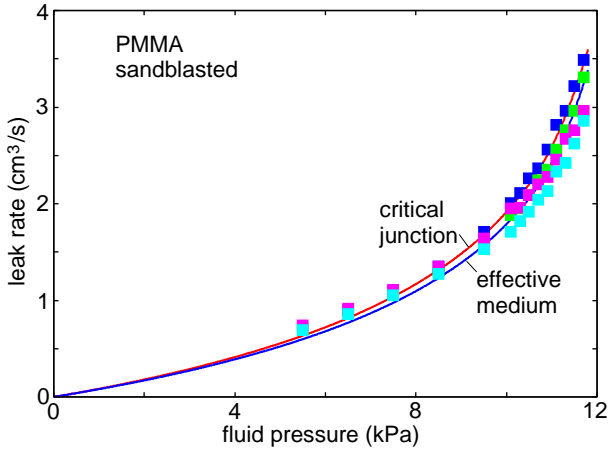}
\caption{\label{without.5}
Fluid leak rate as a function of the fluid pressure difference $\Delta p = p_{\rm a}-p_{\rm b}$. The nominal squeezing pressure
is $p_0 = 11.8 \ {\rm kPa}$. The square symbols are measured data while the solid lines are the theory predictions.
In the calculation we used $\alpha = 0.54$. For sandblasted PMMA with the root-mean-square roughness $34 \ {\rm \mu m}$.} 
\end{figure}

Let us compare the theory to experiment. In Fig. \ref{without.5} we show
the fluid leak rate as a function of the fluid pressure difference $\Delta p = p_{\rm a}-p_{\rm b}$. 
The square symbols are measured data while the solid lines are the theory predictions. Note that the
fluid leak rate rapidly increases when the fluid pressure $\Delta p$ approaches the nominal squeezing
pressure $p_0=11.8 \ {\rm kPa}$. Both the critical junction and effective medium theories predict nearly the same 
pressure dependence of the leak-rate as observed in the experiment.

\begin{figure}
\includegraphics[width=0.45\textwidth,angle=0]{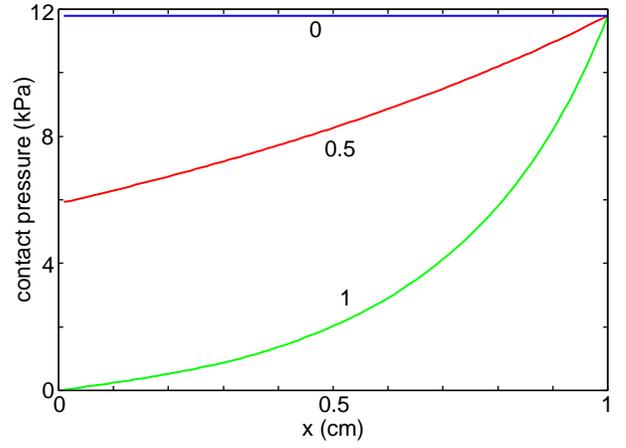}
\caption{\label{x.pcontact}
The contact pressure as a function of the distance $x$ between the high-pressure and low-pressure side. 
The nominal squeezing pressure
is $11.8 \ {\rm kPa}$. Calculations are shown for $(p_{\rm a}-p_{\rm b})/p_0 = 0$, $0.5$ and $1$. 
For sandblasted PMMA with the root-mean-square roughness $34 \ {\rm \mu m}$.} 
\end{figure}

\begin{figure}
\includegraphics[width=0.45\textwidth,angle=0]{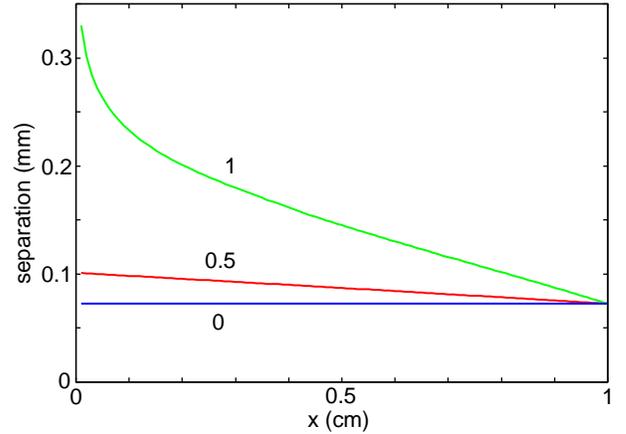}
\caption{\label{x.separation}
The interfacial separation as a function of the distance $x$ between the high-pressure and low-pressure side. 
The nominal squeezing pressure
is $p_0=11.8 \ {\rm kPa}$. Calculations are shown for $(p_{\rm a}-p_{\rm b})/p_0 = 0$, $0.5$ and $1$. 
For sandblasted PMMA with the root-mean-square roughness $34 \ {\rm \mu m}$.} 
\end{figure}

\begin{figure}
\includegraphics[width=0.45\textwidth,angle=0]{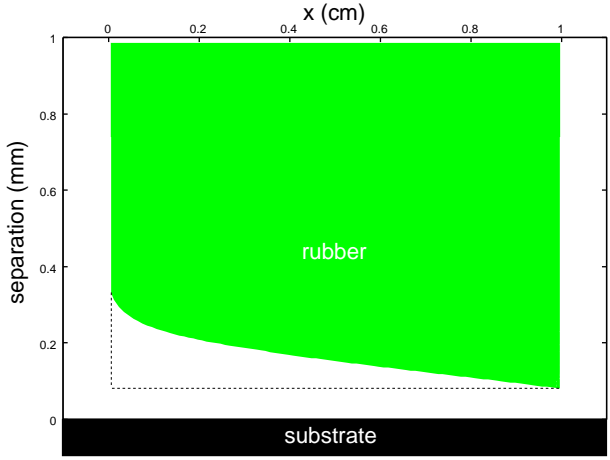}
\caption{\label{rubberProfile}
Shape of the rubber block for $p_{\rm a}-p_{\rm b} = p_0$.
The nominal squeezing pressure
is $p_0=11.8 \ {\rm kPa}$. The dashed line is
the shape of the block when $p_{\rm a}-p_{\rm b} = 0$ 
For sandblasted PMMA with the root-mean-square roughness $34 \ {\rm \mu m}$.} 
\end{figure}

In Fig. \ref{x.pcontact} we show
the calculated contact pressure as a function of the distance $x$ between the high-pressure and low-pressure side. 
Calculations are shown for $(p_{\rm a}-p_{\rm b})/p_0 = 0$, $0.5$ and $1$. 
In Fig. \ref{x.separation} we show
the interfacial separation as a function of the distance $x$ between the high-pressure and low-pressure side. 
In Fig. \ref{rubberProfile} we show for  $p_{\rm a}-p_{\rm b} = p_0$ the deformed rubber block. The dashed
line indicate the rubber block when $p_{\rm a}-p_{\rm b}=0$. In this case the average separation is determined by the 
substrate surface roughness.

\vskip 0.3cm
{\bf 9. Summary and conclusion}

In this paper we have studied fluid squeeze-out from the interface between an elastic solid
with a flat surface and a randomly rough surface of a rigid solid. We have presented a very general
formalism for calculating the (average) interfacial separation as a function of time.
In the theory enters the effective flow conductivity $\sigma_{\rm eff}$. This quantity is a function
of the (local) contact pressure $p_{\rm cont}$. In this paper we have calculated $\sigma_{\rm eff}$
using the so called average-separation and critical-junction theories. An even more accurate method,
based on the Bruggeman effective medium theory, was developed in Ref. \cite{subm}, but this theory
gives results very similar to the critical-junction theory. 
The critical-junction theory and the effective medium theory both consider the flow of fluid in 
interfacial channels and the
possibility (at high enough squeezing pressures)
of trapped fluid at the interface as a result of percolation of the contact area, resulting
in confined regions (islands) of non-contact area filled with fluid. We have shown how this affect the time dependence of the 
interfacial separation. We have shown how the present theory can be used to calculate the leak-rate of static seals
when including the reduction in the contact pressure resulting from the fluid pressure acting
on the solids in the interfacial region. We have presented new experimental data which agree well
with the theory prediction.

\vskip 0.3cm

{\bf Acknowledgments}

This work, as part of the European Science Foundation EUROCORES Program FANAS, was supported from funds 
by the DFG and the EC Sixth Framework Program, under contract N ERAS-CT-2003-980409.

\vskip 0.3cm
{\bf Appendix A}

Consider the elastic contact between two solids with randomly rough surfaces.
The (apparent) relative contact area $A(\zeta)/A_0$ at the magnification $\zeta$
can be obtained using the contact mechanics 
formalism developed elsewhere\cite{PSSR,YP,P1,Bucher,PerssonPRL,earlier},
where the system is studied at different magnifications $\zeta$.
We have\cite{P1,PerssonPRL}

$${A(\zeta)\over A_0} = {1\over (\pi G )^{1/2}}\int_0^{P_0} d\sigma \ {\rm e}^{-\sigma^2/4G} 
= {\rm erf} \left ( P_0 \over 2 G^{1/2} \right )\eqno(A1)$$
where
$$G(\zeta) = {\pi \over 4}\left ({E\over 1-\nu^2}\right )^2 \int_{q_0}^{\zeta q_0} dq q^3 C(q)\eqno(A2)$$
where the surface roughness power spectrum
$$C(q) = {1\over (2\pi)^2} \int d^2x \langle h({\bf x})h({\bf 0})\rangle {\rm e}^{-i{\bf q}\cdot {\bf x}}$$
where $\langle ... \rangle$ stands for ensemble average. 
Here $E$ and $\nu$ are the Young's elastic modulus and the Poisson 
ratio of the rubber. The height profile $h({\bf x})$ of the rough surface can be measured routinely
today on all relevant length scales using optical and stylus experiments.

\begin{figure}
\includegraphics[width=0.35\textwidth,angle=0]{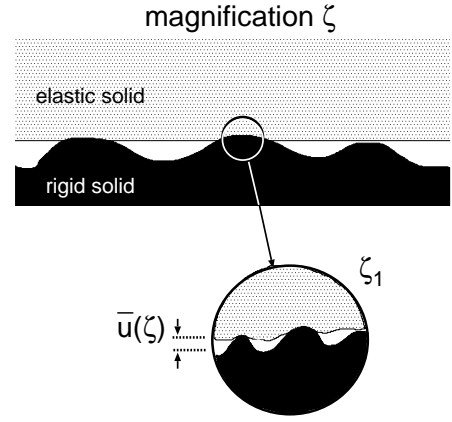}
\caption{\label{ManyMag}
An asperity contact region observed at the magnification $\zeta$. It appears that
complete contact occur in the asperity contact region, but when the magnification is
increasing to the highest (atomic scale) magnification $\zeta_1$, 
it is observed that the solids are actually separated by the average distance $\bar{u}(\zeta)$.
}
\end{figure}

We define $u_1(\zeta)$ to be the (average) height separating the surfaces which appear to come into 
contact when the magnification decreases from $\zeta$ to $\zeta-\Delta \zeta$, where $\Delta \zeta$
is a small (infinitesimal) change in the magnification. 
$u_1(\zeta)$ is a monotonically decreasing
function of $\zeta$, and can be calculated from the average interfacial separation
$\bar u(\zeta)$ and $A(\zeta)$ using
(see Ref.~\cite{YP})
$$u_1(\zeta)=\bar u(\zeta)+\bar u'(\zeta) A(\zeta)/A'(\zeta).\eqno(A3)$$
The quantity $\bar u(\zeta)$ is the average separation between the surfaces in the apparent contact regions
observed at the magnification $\zeta$, see Fig.~\ref{ManyMag}. 
It can be calculated from\cite{YP}
$$\bar{u}(\zeta ) = \surd \pi \int_{\zeta q_0}^{q_1} dq \ q^2C(q) 
w(q,\zeta) $$
$$\times \int_{p(\zeta)}^\infty dp' 
 \ {1 \over p'} e^{-[w(q,\zeta) p'/E^*]^2},\eqno(A4)$$
where $p(\zeta)=p_0A_0/A(\zeta)$ (where $p_{\rm 0} = p_{\rm cont}$ denote the nominal contact pressure)
and
$$w(q,\zeta)=\left (\pi \int_{\zeta q_0}^q dq' \ q'^3 C(q') \right )^{-1/2}.$$

We will now show that as $p_0 \rightarrow 0$, for the values of the magnification $\zeta$
which are most important for the fluid flow between the solids, $u_1 (\zeta) \rightarrow \bar u(\zeta)$. 
This result is physically plausible because at low contact pressures the separation between the walls is
large and the surface roughness should have a very small 
influence on the fluid flow, which therefore can be accurately studied using 
the (average) interfacial separation $\bar u$ for $\zeta \approx 1$.
 
Most of the fluid flow occur in the flow channels which appear
close to the percolation limit where $A(\zeta)/A_0 \approx 0.5$, or, using (A1), 
for $p_0/ G^{1/2}(\zeta) \approx 1$. 
Thus as $p_0\rightarrow 0$ we must have $G(\zeta) \rightarrow 0$ which, using (A2), implies $\zeta \rightarrow 1$.
In fact, using  $G(\zeta)\sim p_0^2$ and (A2) one can easily show that for $\zeta$ close to unity $\zeta -1 \sim p_0^2$.
From (A4) it is easy to show that as $p_0 \rightarrow 0$, the average
separation $\bar u(\zeta)$ will diverge as $\sim - {\rm log} p_0$ while $\bar u'(\zeta)$ diverge as
$-p'(\zeta)/p(\zeta) = A'(\zeta)/A(\zeta)$. Thus the product $\bar u'(\zeta) A'(\zeta)/A(\zeta)$ remains constant
as $p_0\rightarrow 0$. It follows from (A3) that as $p_0\rightarrow 0$, $u_1(\zeta) \rightarrow \bar u(\zeta)$.  

Note that from (A1) and (A2) one can calculate
$${A'(\zeta) \over A(\zeta)} \approx {1\over 4(\zeta - 1)} \sim p_0^{-2}$$
so the slope of the curve $A(\zeta)$ in the relevant $\zeta$-region becomes very high as $p_0\rightarrow 0$.


\begin{thebibliography}{999}
\bibitem{strictly}
Strictly speaking, within fluid continuum mechanics for wetting liquids 
complete fluid squeeze-out never occur. 
We have defined the area of real contact $A$ by assuming that it can be obtained from
the contact pressure $p_{\rm cont}$ using the contact mechanics theory of dry contacts.
\end{thebibliography}
\end{document}